\documentclass[12pt,english]{article}
\usepackage{mathptmx}
\usepackage[T1]{fontenc}
\usepackage[latin9]{inputenc}
\usepackage[a4paper]{geometry}
\usepackage{amsmath}
\usepackage{graphicx}
\usepackage{amssymb}

\makeatletter

\providecommand{\tabularnewline}{\\}

\newcommand {\bek}{\begin{equation}}
\newcommand {\eek}{\end{equation}}
\newcommand {\bea}{\begin{eqnarray}}
\newcommand {\eea}{\end{eqnarray}}

\usepackage{breivik} 

\makeatother

\usepackage{babel}

\begin{document}

\title{An Operational Search and Rescue Model for the Norwegian Sea and
the North Sea}

\author{Øyvind Breivik%
\footnote{Published as Breivik, Ø, and A Allen, 2008: An operational search
and rescue model for the Norwegian Sea and the North Sea, \emph{J
Marine Syst}, \textbf{69}(1-2), 99-113,
doi:10.1016/j.jmarsys.2007.02.010}
\thanks{The Norwegian Meteorological Institute, Alleg 70, NO-5007 Bergen,
Norway%
}%
\thanks{Corresponding author. E-mail: oyvind.breivik@met.no%
} and Arthur A. Allen%
\thanks{U.S. Coast Guard, Office of Search and Rescue, 1082 Shennecossett
Road Groton, CT, USA%
}}
\date{Available online 15 February 2007}
\maketitle
\begin{abstract}
A new operational, ensemble-based search and rescue model for the Norwegian
Sea and the North Sea is presented. The stochastic trajectory model
computes the net motion of a range of search and rescue objects. A
new, robust formulation for the relation between the wind and the
motion of the drifting object (termed the leeway of the object) is
employed. Empirically derived coefficients for 63 categories of search
objects compiled by the US Coast Guard are ingested to estimate the
leeway of the drifting objects. A Monte Carlo technique is employed
to generate an ensemble that accounts for the uncertainties in forcing
fields (wind and current), leeway drift properties, and the initial
position of the search object. The ensemble yields an estimate of
the time-evolving probability density function of the location of
the search object, and its envelope defines the search area. Forcing
fields from the operational oceanic and atmospheric forecast system
of The Norwegian Meteorological Institute are used as input to the
trajectory model. This allows for the first time high-resolution wind
and current fields to be used to forecast search areas up to 60 hours
into the future. A limited set of field exercises show good agreement
between model trajectories, search areas, and observed trajectories
for liferafts and other search objects. Comparison with older methods
shows that search areas expand much more slowly using the new ensemble
method with high resolution forcing fields and the new leeway formulation.
It is found that going to higher-order stochastic trajectory models
will not significantly improve the forecast skill and the rate of
expansion of search areas.
\end{abstract}

\subsection*{Keywords}

Search and rescue, operational forecasting, leeway, stochastic modelling,
trajectory modelling.

\subsection*{Regional terms}

North Atlantic, Norwegian Sea and North Sea.

\section{Introduction}

\label{Sec:intro} The Norwegian Joint Rescue Coordination Centres
(JRCC) handle more than 1500 maritime incidents each year in the Norwegian
Sea and surrounding waters. Of these incidents, a substantial part
involves both search and rescue%
\footnote{Source: the official statistics of the Norwegian RCC, 2001%
} (SAR)%
\footnote{Not to be confused with synthetic aperture radar, also commonly referred
to as SAR.%
}. This was the motivation for developing an operational search and
rescue model that could be initiated with a minimum of information
and that would rapidly return search areas based on prognoses of wind
and surface currents.

Maritime search and rescue is essentially about estimating a search
area by quantifying a number of unknowns (the last known position,
the object type and the wind, sea state, and currents affecting the
object), then compute the evolution of the search area with time and
rapidly deploy search and rescue units (SRU) in the search area. This
puts certain constraints on the model. First, the degrees of freedom
must be limited to allow easy operation. This means that uncertainty
about the last known position and assumptions on the shape of the
object must be tractable for operational users in real time applications.
Second, environmental data (wind and current fields, either prognostic,
observed or climatological) must be available in real time and third,
the model must be fast enough to make it an instrument for operative
search area planning. 

The motion of a drifting object on the sea surface is the net result
of the balance of forces acting on it from the wind, the currents
and the waves. In theory it is possible to compute the trajectory
of an arbitrary drifting object given sufficient information on the
shape and buoyancy of the object, the wind and wave conditions, and
the surface current. In practice, the net motion is difficult to compute
due to the irregular geometry of real-world objects. Simplifications
must be introduced to make the problem tractable, and with these simplifications
errors will also be introduced.

The drift brought about by the wind alone is termed the object's \emph{leeway}.
We follow the definition by \shortciteN{hod95} that {}``The leeway
is the drift associated with wind forces on the exposed above-water
part of the object''%
\footnote{Although the operational model presented here has been given the name
\textsc{Leeway} it computes the net motion brought about both by the
wind and the surface current. %
}. The aerodynamic force from the wind has a drag and a lift component
due to the asymmetry of the overwater structure of the object \cite{ric97}.
The drag is in the relative downwind direction%
\footnote{The relative wind vector is the wind minus the motion of the object;
$\mathbf{W}_{\mathrm{r}}=\mathbf{W}-\mathbf{V}$.%
}, whereas the lift is perpendicular to the relative wind direction
and will cause the object to diverge from the downwind direction.
Likewise, the hydrodynamic force on the submerged part can be decomposed
into a lift and a drag component. The hydrodynamic lift will balance
the lateral (perpendicular to the drift direction) component of the
aerodynamic force and prevent the object from toppling sideways. The
lift gives rise to a significant crosswind leeway component for elongated
objects. This phenomenon is most commonly observed with sailboats
that indeed are designed to sail up against the wind \cite[pp 575--576]{kun90}.
The object's initial orientation relative to the wind (left or right
of downwind) will set the object off along different paths. As the
initial orientation is essentially unpredictable equal probability
must be assigned to the two outcomes.

Another source of uncertainty lies in the wind and current data, modelled
or observed. The fields will always contain errors. Additionally,
fluctuations on a scale smaller than those resolved by the forecast
models or observing systems will always be present. The phenomenon
is often referred to as sub-grid scale in numerical modelling. Similarly,
observing systems will also have temporal and spatial resolution issues,
often referred to as the error of representativeness \cite[p 12]{dal91}.
These unresolved and unobserved small scale fluctuations will affect
the drifting object and must be quantified and taken into account
when estimating its motion. The forces acting on an object on the
sea surface are discussed in Sec~\ref{sec:theory} along with the
simplifications and approximations employed.

It is obvious that there is a large ingredient of chance involved
in the calculation of an object's motion on the sea surface, thus
a probalistic formulation is necessary to tackle the uncertainties
involved. Rather than forecasting the exact trajectory of the object,
a most probable area is sought, i.e., an evolving probability density
function in space. We employ a Monte Carlo technique to compute the
probability density function (interpreted as the search area) for
the location of the object by perturbing the different parameters
that have a bearing on the object's trajectory. The ensemble (Monte
Carlo) trajectory approach is discussed in Sec~\ref{Sec:stochastic trajectory}.

The problem is made even more complicated when we go from modelling
the drift of a known object from a last known position to searching
for a possibly unknown object with scarce information on its last
known position. Here, a geographic area and a time period enter the
suite of uncertainties along with the unknown drift properties of
the object at large. Sec~\ref{sec:operational} describes the implementation
of the operational search model \textsc{Leeway}. 

The \textsc{Leeway} model is part of a suite of oceanic trajectory
models including a ship drift model and a three dimensional oil drift
model. The models are developed and operated by the Norwegian Meteorological
Institute for the operational community (search and rescue and vessel
traffic service and the enviromental protection agency). A description
of the complete trajectory model suite is given by \shortciteN{hac06}.

An earlier Monte Carlo based search model has been successfully used
in maritime search operations since the early seventies when the US
Coast Guard developed its {}``Computer assisted search program''
(CASP). For a review of the developments in this field, see \citeN{fro01}.
The major difference between earlier Monte Carlo based SAR models
and our approach is the quality of the atmospheric and oceanographic
fields (e.g\emph{.,} real-time currents yield lower error variance
than climatological currents typically used in earlier systems) and
the formulations of the leeway coefficients, which are here decomposed
into the more robust downwind and crosswind components rather than
a leeway divergence angle (discussed in Sec~\ref{Sec:experiments})

Sec~\ref{sec:discussion} discusses some results obtained with the
model, the limitations of the theory, the quality of the operational
tool, and possible future extensions.

\section{The forces on a drifting object}

\label{sec:theory}\label{Sec:empirical trajectory}An object on the
sea surface will accelerate as\begin{equation}
(m+m')\frac{d\mathbf{V}}{dt}=\sum\mathbf{F},\label{eq:forces}\end{equation}
 where $\mathbf{V}$ is the object velocity and $\sum\mathbf{F}$
is the sum of the forces. The added mass $m'$ comes from the acceleration
of water particles along the hull of the object (\citeNP{hod98}, \citeNP{mei89}, \citeNP{ric97}).

Life rafts are observed by \shortciteN{hod98} to reach terminal velocity
in approximately 20s under strong wind conditions (20m/s). This implies
that small objects (typically less than 30 m) accelerate very rapidly.
This is confirmed by \shortciteN{fit94} who found the highest cross-correlation
between leeway speed and wind speed at zero lag for 10-minute vector
averaged samples. Infinite acceleration and constant velocity for
the duration of a model time step are thus acceptable simplifications.
The length of the model timestep is then dictated by the temporal
and spatial scale of the forces acting on the object (wind, waves
and currents). These vary over much longer timescales (surface currents,
including tidal motion, and the synoptic weather situation change
appreciably in hours, not minutes).

\subsection{Wind and leeway}

\label{Sec:experiments}We define the leeway (windage) of an object
to be the drift associated with the wind force on the overwater structure
of the object as measured relative to the 10 minute averaged wind
measured at 10 m height (or reduced to this height). This coincides
with the meteorological convention for measuring surface wind and
is consistent with earlier work in this field (\shortciteNP{hod95}).
It is observed that an object moving under the influence of the wind
will diverge to some extent from the downwind direction due to balance
between the hydrodynamic lift and drag of the subsurface area and
the aerodynamic lift and drag of the wind.

\begin{figure}
\begin{centering}
\includegraphics[scale=0.7]{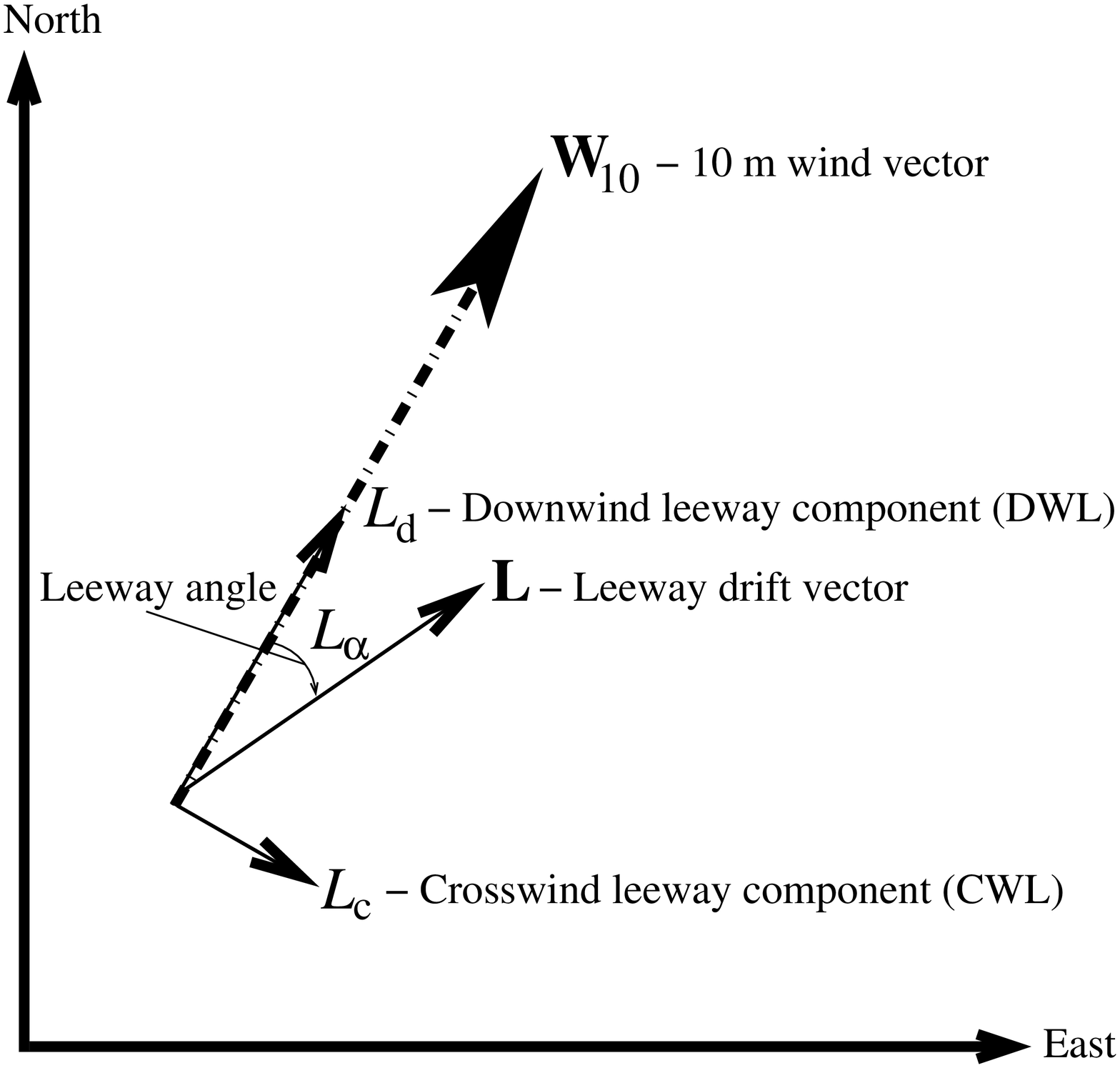}
\par\end{centering}

\caption{\label{Fig:dwlcwl}The \emph{leeway} $\mathbf{L}$ of a drifting object
consists of a \emph{downwind component} (DWL), $L_{\mathrm{d}}$,
and a \emph{crosswind component} (CWL), $L_{\mathrm{c}}$. The angle
between the downwind direction and the leeway drift direction is termed
the \emph{leeway divergence angle}, $L_{\alpha}$ {[}adapted from
Allen and Plourde, 1999{]}.}

\end{figure}

The empirical relation between leeway and wind speed presented here
is based on field work carried out by or compiled from other sources
by the U.S. Coast Guard. The empirical coefficients are summarised
in \shortciteN{all99}. A total of 63 different search leeway categories
has been compiled (discussed further in Sec~\ref{Sec:taxonomy},
see also Table~\ref{Tab:taxonomy}). Leeway field experiments endeavour
to determine the relation between the wind speed and the leeway speed
and divergence angle, $L$ and $L_{\alpha}$, or more robustly, the
downwind (DWL) and crosswind (CWL) leeway components, $L_{\mathrm{d}}$
and $L_{\mathrm{c}}$ (see Fig \ref{Fig:dwlcwl}). The latter parameters
are more stable to directional fluctuations at low wind speed and
are therefore preferred. To achieve this, it is necessary to combine
wind measurements from a nearby buoy with GPS measurements of the
motion of the object (over 10 minute intervals) and current measurements
of the slippage (the object's motion relative to the current at a
certain level, discussed below). The empirical coefficients for leeway
speed and the divergence angle listed in \shortciteN{all99} were
converted by \shortciteN{all05} into downwind and crosswind components
of leeway as functions of the 10 m wind speed. The decomposition into
downwind and crosswind coefficients are implemented for the first
time in the operational model described in this work.

\begin{figure}
\includegraphics{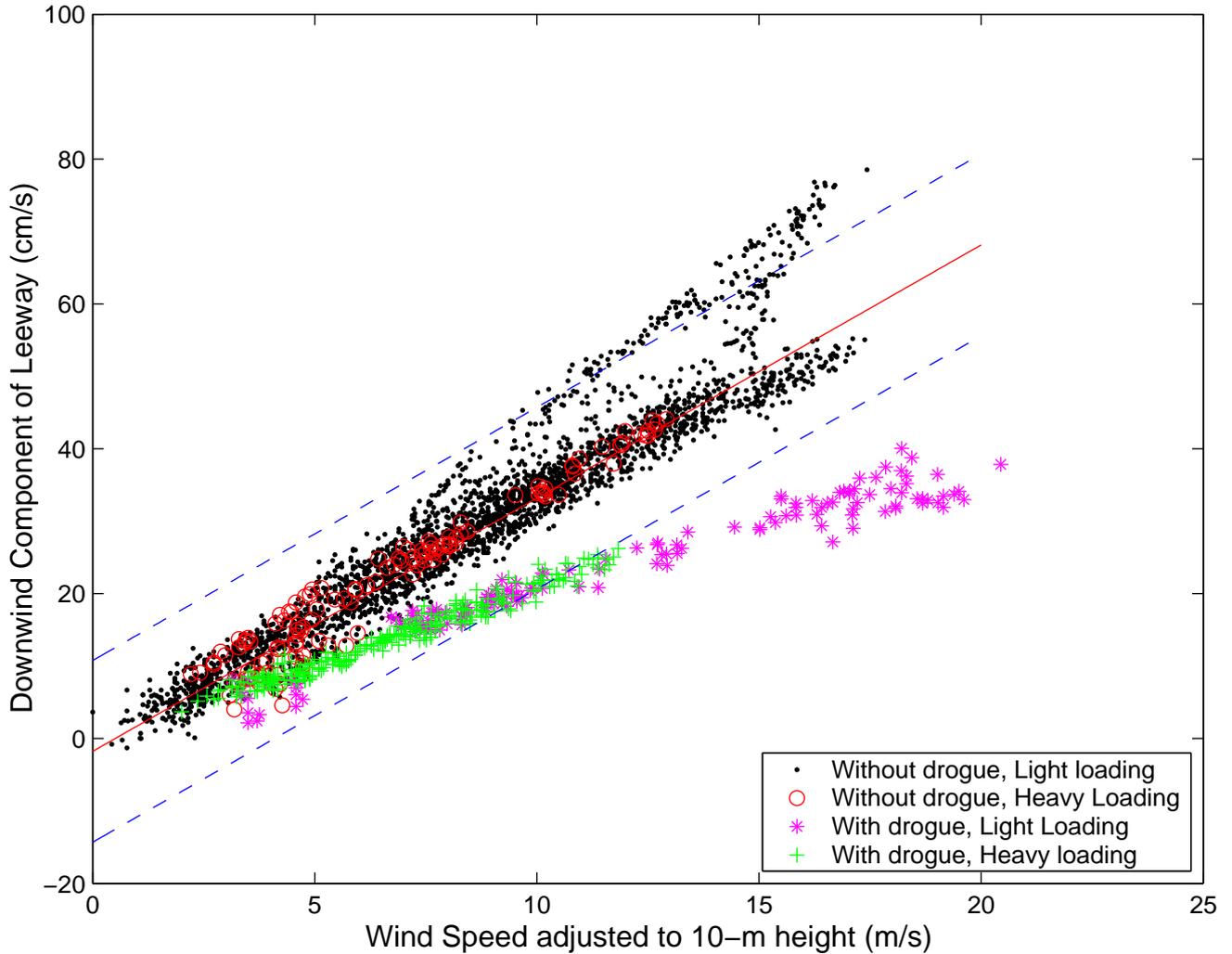}

\caption{\label{Fig:dwl}Measured downwind component of leeway, $L_{\mathrm{d}}$
of maritime life rafts with deep ballast systems, canopy, and capacity
4--6 persons relative to the 10~m wind speed, $W_{10}$. The leeway
measurements are wind only, the wind-induced drift is subtracted from
the measurements using surface current measurements. The linear regression
$\widehat{L}_{\mathrm{d}}$ and its 95\% confidence limits are also
indicated. Note the influence of the drogue (sea anchor, marked with
asterisks in pink and green) on the leeway. The experimental variance
is also a function of the wind speed (hetereoscedasticity, i.e., the
spread is greater for higher wind speeds). {[}from Allen and Plourde,
1999{]}.}

\end{figure}

\begin{figure}
\includegraphics{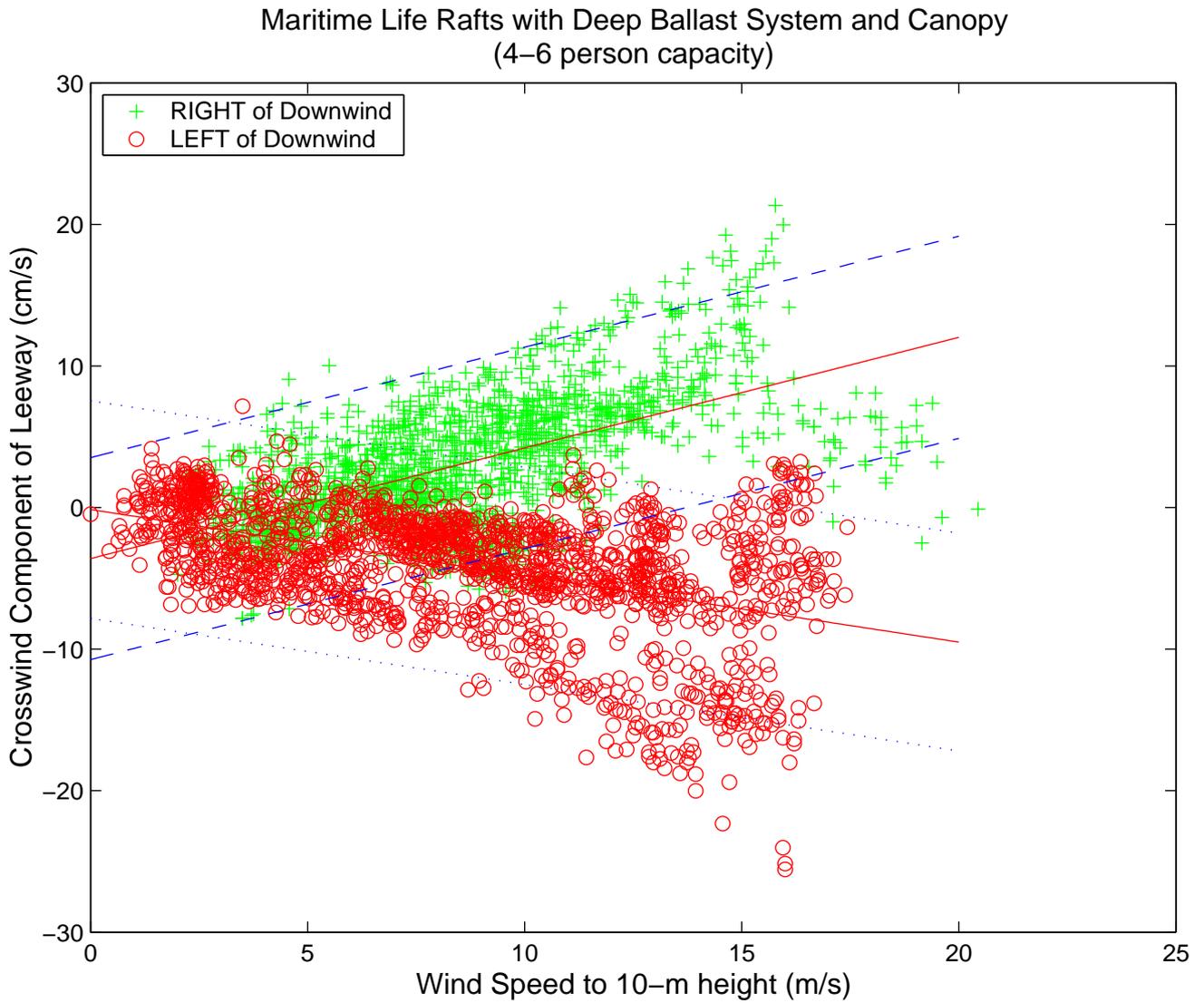}

\caption{\label{Fig:cwl}Same as Fig~\ref{Fig:dwl} for the crosswind component
of leeway. The leeway measurements are wind only, the wind-induced
drift is subtracted from the measurements using surface current measurements.
Drifting objects can move at a relative angle left (red) or right
of downwind (green) (from Allen and Plourde, 1999).}

\end{figure}

The experimental data shown in Fig~\ref{Fig:dwl} suggest an almost
linear relationship between wind speed $W_{10}$ and the downwind
component of the leeway. This linear relationship is also observed
by other workers (see \citeNP{ric97}; \citeNP{hod98}). The downwind
and crosswind leeway measurements described in Figs~\ref{Fig:dwl}
and~\ref{Fig:cwl} are corrected for wind-induced drift by measuring
the slippage.

A linear regression with least squares best fit coefficients is now
computed to relate $L_{\mathrm{d}}$ to the local wind speed,\begin{equation}
\widehat{L_{\mathrm{d}}}=a_{\mathrm{d}}W_{10}+b_{\mathrm{d}}.\label{eq:dwl}\end{equation}
 Here, $a_{\mathrm{d}}$ and $b_{\mathrm{d}}$ are regression coefficients
determined from the experimental data and $\widehat{L_{\mathrm{d}}}$
is the estimated best fit DWL. An analogous derivation applies to
the crosswind component of the leeway, $\widehat{L_{\mathrm{c}}}$.
The dataset is divided into left-drifting and right-drifting observations,
depending on whether the object was observed to bear to the left or
to the right of the local downwind direction (see Fig \ref{Fig:cwl}).
Note that the regression coefficients are similar but not identical
for the left and right drifting objects. The data suggest that left
and right drifting observations are almost equally probable, hence
in the following we assume a 50/50 distribution.

The 10 m wind fields used in the operational ensemble trajectory model
described later are taken from the High Resolution Limited Area Model
(HIRLAM), the numerical weather prediction model of The Norwegian
Meteorological Institute. The model has a horizontal resolution of
approximately 20 km on a rotated spherical grid. The model and assimilation
is run four times daily (00, 06, 12, 18 UTC). Each model integration
extends to +60 hours. The current version of the HIRLAM model is described
by \shortciteN{und02}. Open boundary conditions are taken from the
Integrated Forecast System (IFS) of The European Centre for Medium-Range
Weather Forecasts (ECMWF).

\subsection{Current and slippage}

The current vectors used in this study are taken at 0.5 m depth. This
coincides with the draft of typical SAR objects and roughly with the
depth at which current meters are mounted in leeway field experiments.
The slippage of an object is defined as its motion relative to the
ambient current at a certain depth comparable to the draft of the
object (set to 0.5 m in our case). In the absence of wind, the object
is assumed to move with the local surface current, $\mathbf{u}_{\mathrm{w}}$,
i.e., no slippage. The object is further assumed to adjust its motion
instantaneously when the current changes (infinite acceleration, discussed
above).

Surface current fields in the operational setup of the \textsc{Leeway}
model described in Sec~\ref{sec:operational} are taken from the
operational 3D baroclinic ocean model of The Norwegian Meteorological
Institute. The model is a modified version of the Princeton Ocean
Model (POM) with 4 km horizontal resolution on a polar stereographic
Arakawa C grid. The model solves numerically the primitive equations
of motion (with Boussinesq and hydrostatic approximations) along with
equations for the conservation of heat and salt. A Mellor-Yamada turbulence
closure scheme is employed for the vertical eddy viscosity \cite{mel82}.
The model is driven by atmospheric forcing (10 m wind, heat flux,
and atmospheric pressure) from an operational atmospheric model (see
above) and all major tidal constituents on the open boundary. The
original model formulation is described by \citeN{blu87}. 

The vertical dimension is resolved by 21 unevenly spaced (higher resolution
near the surface and the bottom) layers of terrain-following $\sigma$-coordinates.
The uppermost layer of the model is $0.001H$, where $H$ is the depth.
Coastal waters are thus well resolved vertically (e.g., the uppermost
layer is 0.3 m below the sea surface in 300 m deep waters). For our
purposes current vectors have been interpolated to 0.5 m below the
sea surface for consistency with the field experiments carried out
to establish leeway coefficients.

Climatological river runoff is included in the model. The river runoff
has a strong seasonal signal and makes an important contribution to
the coastal circulation.

\begin{figure}
\includegraphics[scale=0.45,angle=90]{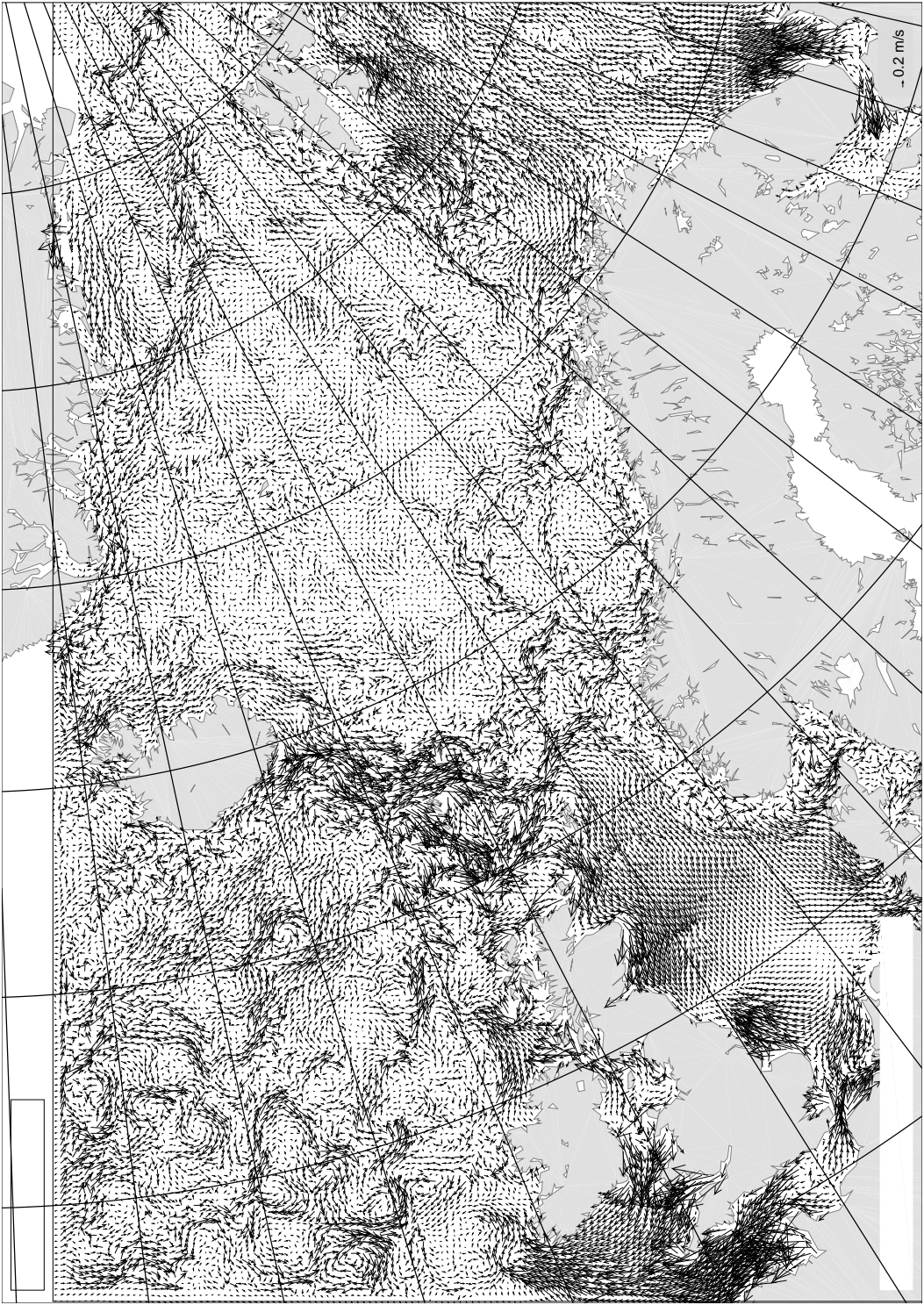}

\caption{\label{Fig:ormen}Modified Princeton Ocean Model domain operated by
The Norwegian Meteorological Institute. Current vectors at 0.5 m below
sea surface are shown. The model has a horizontal resolution of 4
km (every fourth vector shown, $5^{\circ}\ \,$graticule). The ocean
model is forced with 20 km resolution 10 m winds from the HIRLAM numerical
weather prediction model. All major tidal constituents are prescribed
on the boundary. Daily prognoses out to +60 h are initiated at 00
UTC. The operational setup of the \textsc{leeway} model is confined
to the ocean model grid. A seven day archive allows trajectory calculations
to be inititated several days after the event occurred. }

\end{figure}

The model and the operational setup at the Norwegian Meteorological
Institute is described in detail in \citeN{eng95} and \citeN{eng01}.
The model grid is outlined in Fig~\ref{Fig:ormen}.

\subsection{Wave effects ignored}

The Stokes drift is a downwave drift induced by the orbital motion
that water particles undergo under the influence of a wave field.
These particle orbits are not closed and a Lagrangian drift is set
up. This drift is confined to a narrow layer next to the sea surface
\cite[pp 223--225]{kun90}. It is well known that the Stokes drift
can be a dominant factor in the advection of suspended material and
objects on the sea surface.

In the field campaigns conducted to estimate leeway coefficients (reviewed
in \shortciteNP{all99}), the wind and the near-surface current were
the only measured quantities apart from the object's motion. The Stokes
drift is predominantly down-wind and is difficult to separate from
the direct wind effect on the object after the ambient current has
been subtracted because it is a Lagrangian effect invisible to the
Eulerian current measurements that were collected in the field experiments.
Thus, even though the Stokes drift is readily available from operational
wave models such as WAM (see \shortciteNP{has88}), it is necessary
to leave it out as it is assumed to be already present in the empirical
leeway coefficients.

The Stokes drift can also significantly modify the upper layer velocity
through interaction with a sheared near-surface current. This interaction
generates Langmuir circulations, i.e. vortices parallel to the wind
direction (see \shortciteNP{sky95} and \shortciteNP{car05}). No
attempt has been made at estimating the wave-current effect on the
near-surface currents as this is numerically very demanding. It requires
an estimate of the Stokes drift derived from a wave forecasting model
which in turn must be added to the momentum equations of the ocean
model. However, it is clear that this effect may become important,
particularly in high seas and a strong vertically sheared current.

Wavelengths comparable to the object dimension are most efficient
in transferring energy to the drifting object \cite{hod98}. The wave
force in this regime ($\lambda\sim D$) for a \emph{fixed} object
simplifies to \begin{equation}
\mathbf{F}_{\mathrm{wave}}=\frac{1}{4}\rho ga^{2}D.\label{eq:waveforce}\end{equation}
 Here, $a$ represents the amplitude of the energy found in a narrow
spectral band covering the dimension $\lambda=D$ \cite{mei89}. Note
that the energy transmitted to a \emph{drifting} object is even smaller
than in the case of a fixed object. We assume that damping and excitation
is negligible for small objects (less than 30~m length) with no way
on, even in a well developed sea (see \shortciteNP{kom84}, \shortciteNP{sor98}, and \shortciteNP{wmo88}.
Wave effects are therefore ignored for all objects treated in this
work. It can further be argued that as waves are wind-driven (ignoring
the negligible effect of swell on smaller objects), the wave force
will be aligned with the wind. This will lead to a small downwind
force which will be difficult to disentangle from the field data for
the empirical leeway coefficients as no direct wave measurements have
been performed \cite{all99}.

\subsection{Total motion}

Following the assumptions and simplifications above, the trajectory
model simplifies to calculating the arc traced by the leeway vector
superposed on the surface current,

\[
\mathbf{x}(t)-\mathbf{x}_{0}=\int_{0}^{t}\mathbf{V}(t')\, dt'=\int_{0}^{t}\left[\mathbf{L}(t')+\mathbf{u}_{\mathrm{w}}(t')\right]\, dt'.\]
A second order Runge-Kutta scheme is used for the computation of trajectories
on a sphere.

\section{The operational ensemble trajectory model}

\label{sec:operational}In searching for drifting objects on the sea
surface, we are faced with the challenge of quantifying a large number
of unknowns and to estimate a most probable search area in light of
these uncertainties. A search operation attempts to maximize the probability
of success ($POS$), which in turn depends on the probability of detection
($POD$) and the probability of containment ($POC$),\[
POS=POD\times POC.\]
 $POC$ is the a priori search area, i.e\emph{.} the area that most
likely contains the drifting object. \textsc{Leeway}, the operational
SAR model described here yields search areas ($POC$) and does not
involve the probability of detection, which depends on the resources
available, the visibility in the area and other external factors.
It is the task of the rescuers to take the information from $POC$
and apply their resources ($POD$) to maximize $POS$. In addition
to perturbing the leeway coefficients and the forcing fields, an operational
model must perturb the initial position and the initial release of
the object to account for the uncertainty in last known position.

\label{Sec:stochastic trajectory}The linear leeway relation described
in Sec~\ref{Sec:experiments} can be superposed on surface current
data (prognoses, measurements, or climatology, depending on what is
available) to estimate the trajectory of a drifting object. However,
as was mentioned in Section~\ref{Sec:intro}, there are several sources
of uncertainty and errors that will cause the true and modelled trajectories
to diverge with time. Consequently, an evolving probability density
function in the two lateral dimensions (longitude and latitude) is
sought. The approach chosen here is to estimate the errors and uncertainties
and to perform a Monte Carlo integration by rerunning the trajectory
model multiple times with all relevant parameters perturbed to generate
an ensemble.

To motivate the use of the Monte Carlo technique, we start by assuming
that the position of a drifting object behaves as a Markov process
or first order autoregressive process,\begin{equation}
p(\mathbf{x}_{t+1}|\,\mathbf{x}_{t},\mathbf{x}_{t-1},\mathbf{x}_{t-2},\ldots,\mathbf{x}_{1})=p(\mathbf{x}_{t+1}|\,\mathbf{x}_{t}),\label{Eq:Markov}\end{equation}
 i.e\emph{.}, the probability density function of the future state
is only dependent on the current state and not on the particular way
that the model system arrived at this state (see \citeNP{wil95}, p 287 or \citeNP{pri81}, p 117).
Here, $\mathbf{x}$ represents the state of the system (in our case
the {}``state'' is simply the location of the drifting object and
whether it is drifting or stranded). Let $\mathbf{V}$ represent the
(possibly nonlinear) function for the displacement of the object under
the influence of external forces. Embedded in $\mathbf{V}$ is the
external forcing (wind and current) as well as the drift properties
of our particular object. The random perturbations $d\boldsymbol{\epsilon}$
have a known covariance and zero mean. These represent the uncertainties
in drift properties and forcing (wind and current). Under the assumption
that the perturbations behave as a Markov process, the trajectory
evolves as 

\begin{equation}
d\mathbf{x}=\mathbf{V}(\mathbf{x},t)dt+d\boldsymbol{\epsilon}.\label{SDE}\end{equation}
The ensemble consists of $\mathcal{O}(N)$ samples drawn from the
stochastic differential equation~(\ref{SDE}).

\subsection{Leeway coefficient perturbations}

The orientation (left and right of wind) of the ensemble members is
distributed equally and remains fixed throughout the integration.
This means that a member will not {}``jibe'', i.e., once started
to the left of downwind it will continue on that tack throughout the
simulation.

The perturbations in leeway coefficients are supposed to cater for
the properties unique to the object (e.g., is this a less than normally
loaded life-raft, or does the raft sag?). It is important to note
that the perturbations in leeway properties remain fixed throughout
the simulation.

\begin{figure}
\includegraphics{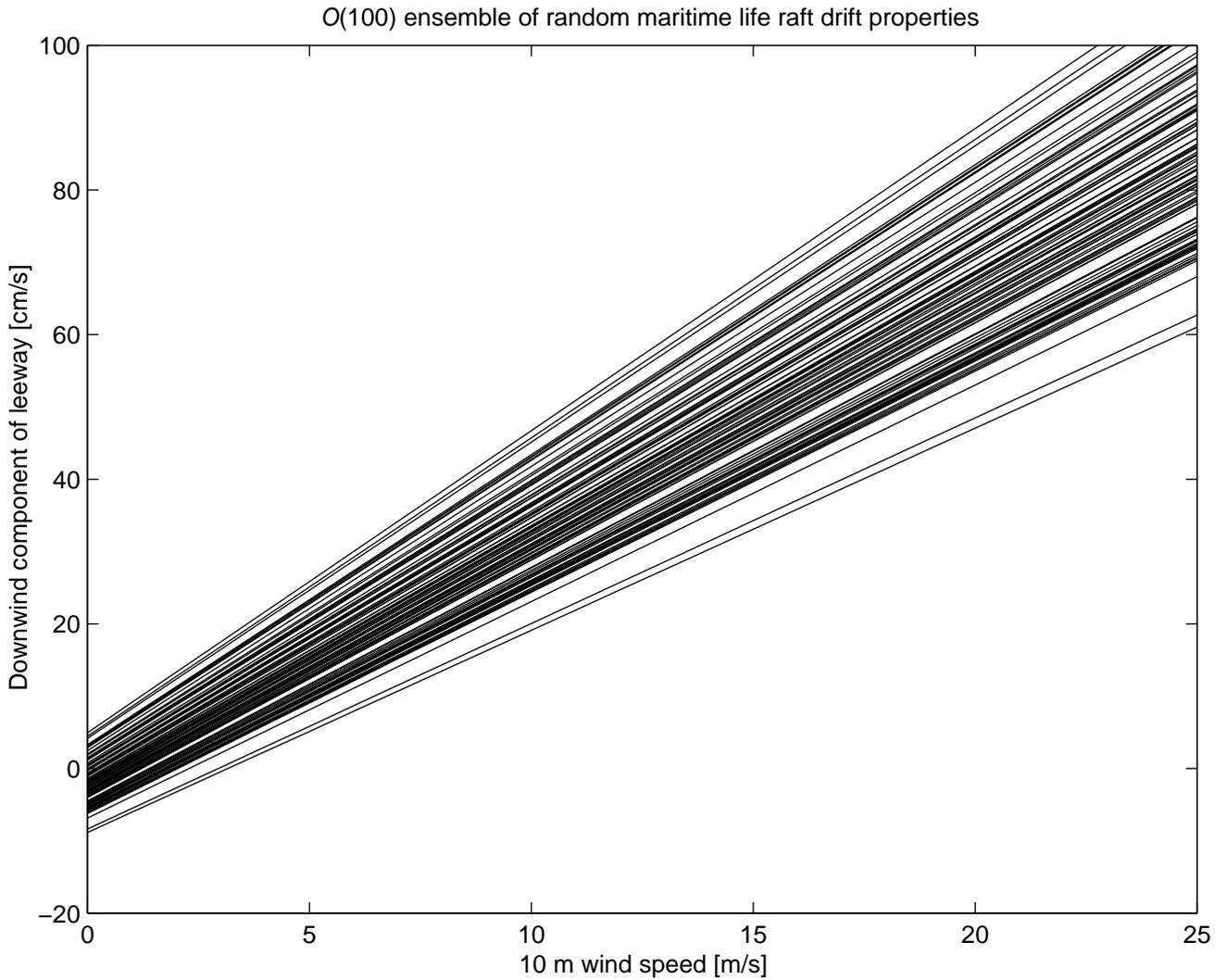}

\caption{\label{Fig:dwltheory}$\mathcal{O}(100)$ random perturbations of
the downwind leeway component of the maritime life raft found in Fig~\ref{Fig:dwl}.
Both the slope and the offset are perturbed to allow the ensemble
to reproduce the heteroscedastic spread about the best fit linear
regression, $L_{n}=(a+\epsilon_{n}/20)W_{10}+(b+\epsilon_{n}/2)$.}

\end{figure}

It is clear from the experimental data compiled by \citeN{all99}
that the variance increases with wind speed, i.e\emph{.}, the dataset
is slightly heteroscedastic (see Figs~\ref{Fig:dwl} and~\ref{Fig:cwl}).
This should be accounted for, and to recreate the spread about the
regression (\ref{eq:dwl}) both the slope and the offset of the regression
line of ensemble members $n=1,\ldots,N$ is adjusted by adding a common
perturbation $\epsilon_{n}$ drawn from a normal distribution $N(0,\sigma)$,

\begin{equation}
a_{n}=a+\epsilon_{n}/20,\label{eq:apert}\end{equation}
\begin{equation}
b_{n}=b+\epsilon_{n}/2.\label{eq:bpert}\end{equation}
 At $W_{10}=10$m/s, equal perturbations are added to slope and offset,
$L=(a+\epsilon_{n}/20)W_{10}+(b+\epsilon_{n}/2)$. At lower wind speeds
the bulk of the perturbation is contributed by the offset ($b+\epsilon_{n}/2$),
whereas at higher wind speeds the slope adds most. Fig~\ref{Fig:dwltheory}
illustrates an ensemble of simulated regression lines drawn from Eqs~(\ref{eq:apert})
and (\ref{eq:bpert}).

\subsection{Wind and current perturbations}

\subsubsection*{Ignoring higher-order dispersion}

Higher-order stochastic particle dispersion models and their applicability
to surface drifting objects with little or no leeway have been studied
extensively by, among others, \citeN{gri96} and \citeN{ber02}. The
models are all Markov processes of different orders. The successor
to the {}``random walk'' or Markov-0 model presently employed by
our trajectory model is the {}``random flight'' formulation. This
transport model allows fluctuations to slowly evolve. The integral
time scale $\mathcal{T}$ is a measure of the memory of the perturbations.
It is related to the autocorrelation of the velocity field through
\[
\mathcal{T}=\int_{0}^{\infty}R(\tau)\, d\tau.\]
An exponential autocorrelation function is normally assumed,\[
R(\tau)=e^{-\tau/T}\]
\cite{gri96}. Upper ocean current fluctuations are correlated and
$\mathcal{\mathcal{T}\sim}2$ days \cite{gri96,pou01}. 

SAR objects are influenced by both the wind and the surface currents.
Ideally, correlated fluctuations in the wind speed and direction should
also be accounted for. \citeN{abd02} found that wind gusts obey a
first order auto-regressive formulation similar to the random flight
formulation for surface drifters. A coefficient of approximately 0.9
is found to be appropriate to add gusts to six-hourly winds to force
a wave model. 

A random flight formulation based on estimates of the integral time
scale for both the surface current fluctuations and the wind fluctuations
would represent the next level of sophistication in our ensemble trajectory
model. This approach has been tested with good results for passive
drifters with negligible leeway by \citeN{ull06}. The drifter trajectories
were estimated using currents from a chain of high-frequency radars.
Estimates of the integral time scale and the upper ocean diffusion
were available from the HF radar data. However, estimating the integral
time scale of surface current fluctuations and wind fluctuations in
different geographic areas and for different seasons can be difficult.
As we have developed an operational model whose primary task is to
estimate the location and magnitude of search areas for an array of
search objects, we argue that it is better to keep the model simple
pending reliable estimates of the turbulent time scales of the upper
ocean and the atmospheric boundary layer.

However, the most compelling argument for retaining a simple stochastic
model is the enormous dispersion caused by the experimental error
variance of the leeway properties of SAR objects with appreciable
windage (which includes all typical SAR objects, even persons in water).
As perturbations to leeway coefficients must be constant in time for
each ensemble member (otherwise the members representing drifting
objects would have drift properties that changed with time, which
is clearly unrealistic), these contribute two orders of magnitude
more to the dispersion than the time-varying wind and current perturbations.
Model tests with wind standard deviation 2.6m/s and current standard
deviation 0.25m/s yield almost the same ensemble spread (less than
2\% difference after $\mathcal{O}(250)$ timesteps, not shown) as
simulations where the wind and current perturbations were turned off.
Thus wind and current induced dispersion is swamped by the dispersion
caused by the perturbations to the leeway coefficients. Obviously,
regions with strong current shear or sharp atmospheric fronts will
cause the ensemble to diverge more rapidly than under stable and homogeneous
conditions. In general, though, it is reasonable to assume that wind
and current perturbations are of secondary importance and that a random
flight dispersion model will not have discernible impact on the rate
of expansion of the search area.

\subsubsection*{A random walk wind perturbation model}

\label{sec:fields}Consequently, perturbations $\mathbf{u}'$ of the
wind field are assumed to follow a circular normal distribution,\begin{equation}
\mathbf{u}'_{n}\equiv(u_{n}',v_{n}')\in N(0,\sigma_{W}),\label{eq:nwind}\end{equation}
\begin{equation}
W_{n}\equiv\left\Vert \mathbf{W}_{10}+\mathbf{u}_{n}'\right\Vert _{.}\label{eq:pwind}\end{equation}
 The ensemble of leeway components becomes\begin{equation}
L_{n}=a_{n}W_{n}+b_{n},\: n=1,\ldots,N,\label{eq:lpert}\end{equation}
 Again, identical derivations hold for the downwind and crosswind
equations, and subscripts to distinguish the two are left out for
brevity. The perturbations $\mathbf{u}_{n}'$ are uncorrelated in
time. Sub-grid scale fluctuations appear as errors in comparisons
between model prognoses and observations. These fluctuations must
be accounted for as the sub-grid scale affects the drifting object.
The zeroth order approach to adding sub-grid scale effects is to use
uncorrelated perturbations (random walk). A possible extension to
allow time-coherent (random flight) fluctuations is discussed in Sec~\ref{sec:discussion}.

Based on wind observations and the 12 h prognosis at Ocean Weather
Station Mike ($66^{\circ}$N, $002^{\circ}$E), we have estimated
the 12 h forecast RMS error to be $\sigma_{W}=2.6$m/s in both the
east and the north components (not shown) of the wind vector.

\subsection{Release time and initial position}

\begin{figure}
\includegraphics[scale=1.2]{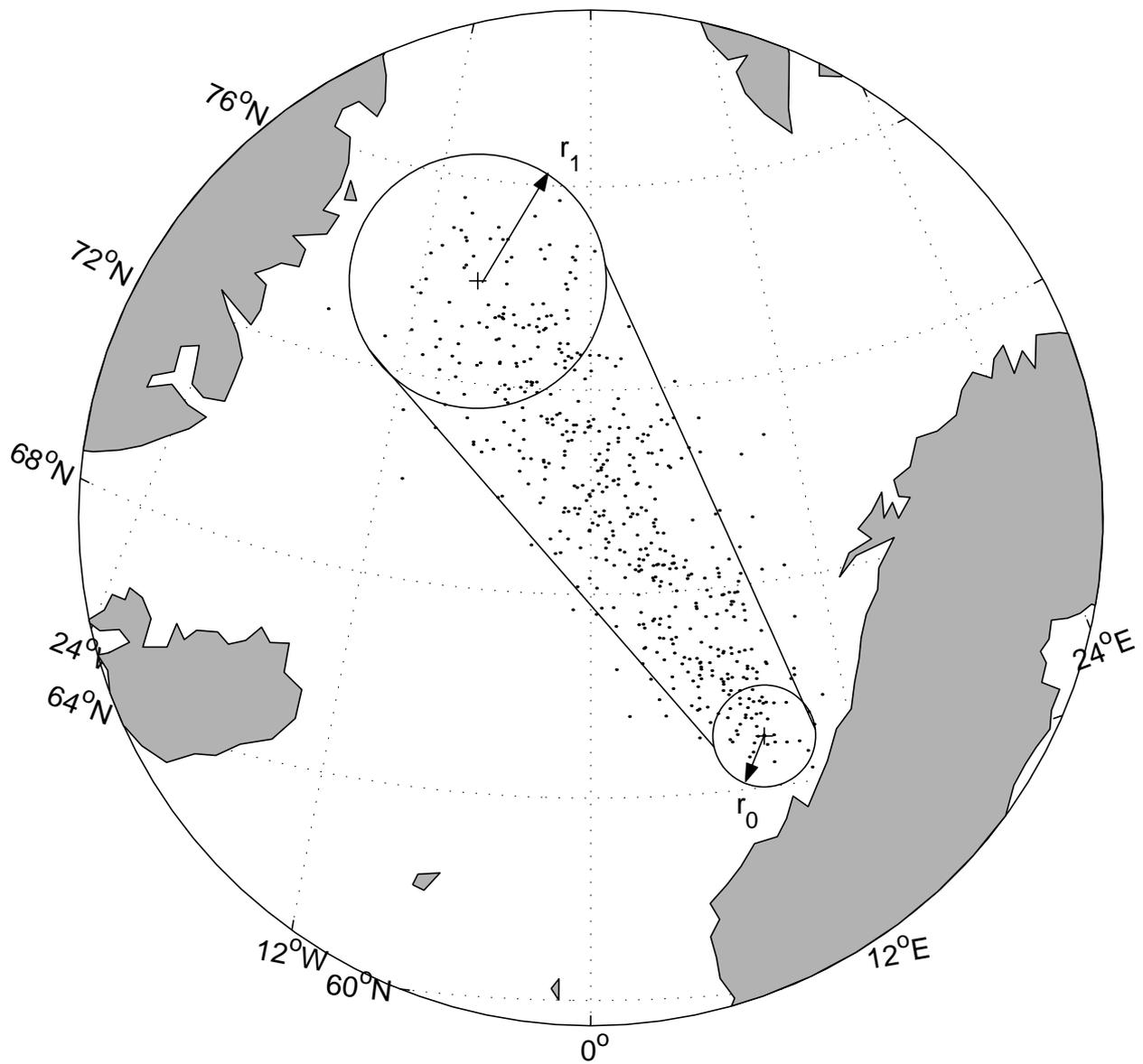}

\caption{\label{Fig:cone}Seeding the ensemble. The initial distribution of
ensemble members in space is defined by two circles with radii $r_{0}$,
and $r_{1}$, centred on positions $(\phi_{0},\lambda_{0})$ and $(\phi_{1},\lambda_{1})$,
respectively. Members (marked as black particles) are seeded randomly
in the area outlined by the two great circle arcs (representing $2\sigma$,
which is why some particles lie outside the arcs). Members are released
continuously between $t_{0}$ and $t_{1}$, the first near position
0, the last near position 1. This allows us to define accidents where
the time of the accident is uncertain while simultaneously assuming
that the vessel is under way. Simpler scenarios can be defined by
setting start and end positions equal og reducing the time period
to a point release.}

\end{figure}

The first task in a search is to determine the object's initial position
in time and space, i.e\emph{.}, the position and point in time where
the object started drifting with no way on. If the last known position
(LKP) is assumed to be rather precise (e.g\emph{.}, a distress call
is received from a ship with a GPS unit), a small radius of uncertainty
may be assigned to this position and consequently all ensemble members
will be released simultaneously and within short distance of each
other. In the other extreme, take a situation where little is known
about the time and location of the accident. Then a wide radius and
a long release period must be used. This will result in a large cloud
of candidate positions at later times. The various members of the
ensemble will endure very different fates due to the spatial and temporal
variation in current and wind fields. It is obvious that the choice
of initial distribution of ensemble members will greatly affect the
future search area. Eight degrees of freedom are available for defining
the initial distribution in the current implementation of \textsc{Leeway}:
\begin{itemize}
\item Date-time, $t_{0},$ position, $\mathbf{x}_{0}$ (latitude and longitude),
and radius of uncertainty, $r_{0},$ of the earliest possible time
of accident
\item Date-time, $t_{1},$ position $\mathbf{x}_{1},$ and radius of uncertainty,
$r_{1},$ of the latest possible time of accident
\end{itemize}
The ensemble size $N$ is $\mathcal{O}(500)$. All initial positions
are drawn from a circular normal distribution with standard deviation
$\sigma=r/2$, where the radius $r$ is varied linearly between $r_{0}$
and $r_{1}$. This approach is flexible, it allows on one hand for
point release in space and time and in the other extreme one can seed
particles continuously in a {}``cone'' shaped area with one radius
of uncertainty in one end and another radius in the other end of the
seeding area. Fig~\ref{Fig:cone} gives an example of a general,
cone shaped initial distribution. By setting $r=2\sigma$, approximately
86\% of the particles are released within the radius.

\subsection{Object taxonomy}

\label{Sec:taxonomy}The drift properties of typical maritime search
objects have been studied extensively since World War II. In Table~\ref{Tab:taxonomy}
we present only a truncated list of the relevant SAR objects discussed
by \citeN{all99}. Some categories are further divided into subclasses.
Generic classes are created by uniting different datasets and increasing
the experimental variance accordingly. Note that the quality of the
experiments, and thus the error variances, varies widely, depending
on the equipment and methods used to conduct the experiments.

\begin{figure}
\includegraphics[scale=0.9]{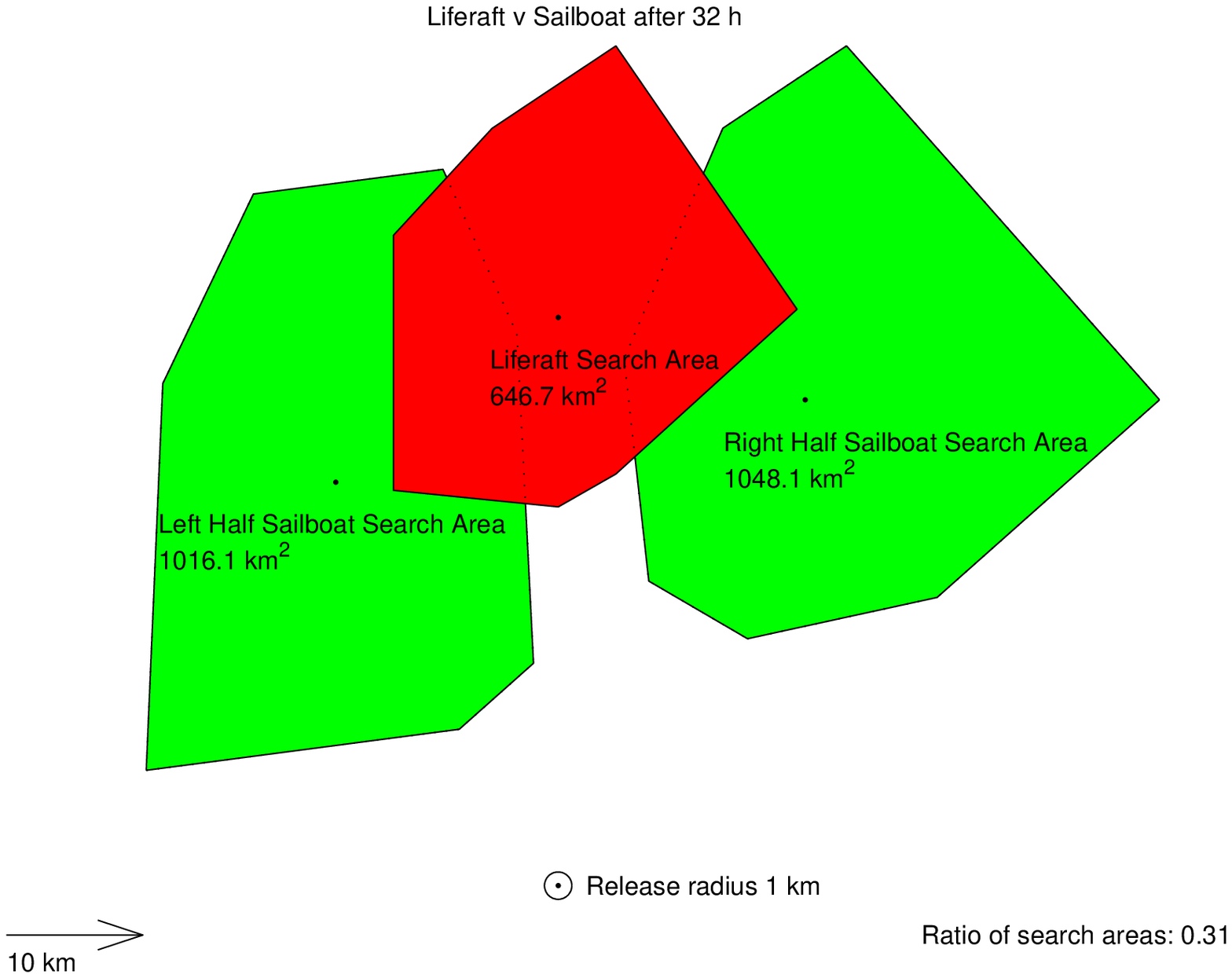}

\caption{\label{Fig:raft_vs_sailboat}}
The simultaneous search areas of a modern life raft with deep ballast
system and a sailboat at time $t=+32\textrm{ hours}$ are shown as
polygons. The average wind speed was 10m/s in this simulation. The
polygons are convex hulls that encompass the individual ensemble members
(not shown). All ensemble members representing both search objects
were released simultaneously in the same radius ($r_{0}=r_{1}=$1~km).
The sailboat search area is significantly larger than the liferaft
search area (left and right lightly shaded polygons). This is in part
due to higher divergence which causes the search area to split in
two after a while. The liferaft search area (dark grey, centre), on
the other hand, is still contiguous as modern liferafts have only
a small crosswind component and thus diverge little from the downwind
direction. This causes the two halves of the ensemble to overlap by
approximately 50\%. The smaller search area of the liferaft is also
a consequence of the improved experimental techniques in the study
of deep ballast liferafts.
\end{figure}

Fig~\ref{Fig:raft_vs_sailboat} illustrates qualitatively the difference
in drift properties between typical search objects, in this case a
life raft and a sail boat, and how this leads to very different search
areas as time progresses. Notice the significant leeway divergence
angle of the sailboat and how this produces two disjoint search areas,
depending on whether the boat drifts to the left or the right of the
wind. The life raft, on the other hand, moves somewhat faster and
with much less divergence, making the overall search area smaller
through overlap. Fig~\ref{Fig:raft_vs_sailboat} also demonstrates
the importance of redoing leeway field work using modern techniques;
the smaller search area of the life raft search area is in part a
consequence of the lower experimental variance achieved in using modern
field experiments.

\section{Discussion and outlook}

\label{sec:discussion}

\subsection{Drift exercises}

\label{sec:eval}Controlled experiments to evaluate the performance
of the operational model are lacking. However, a few incidents have
been reported and a few exercises conducted that yield some information
about the forecast skill of the model. We describe here three cases
reported to us by the Norwegian Rescue Co-ordination Centres and one
specifically modelled. We stress that these cases do not constitute
a model evaluation and serve merely as illustrations of the ability
of the model to reproduce the observed trajectories.

\subsubsection*{Benthic lander}

\begin{figure}
\includegraphics{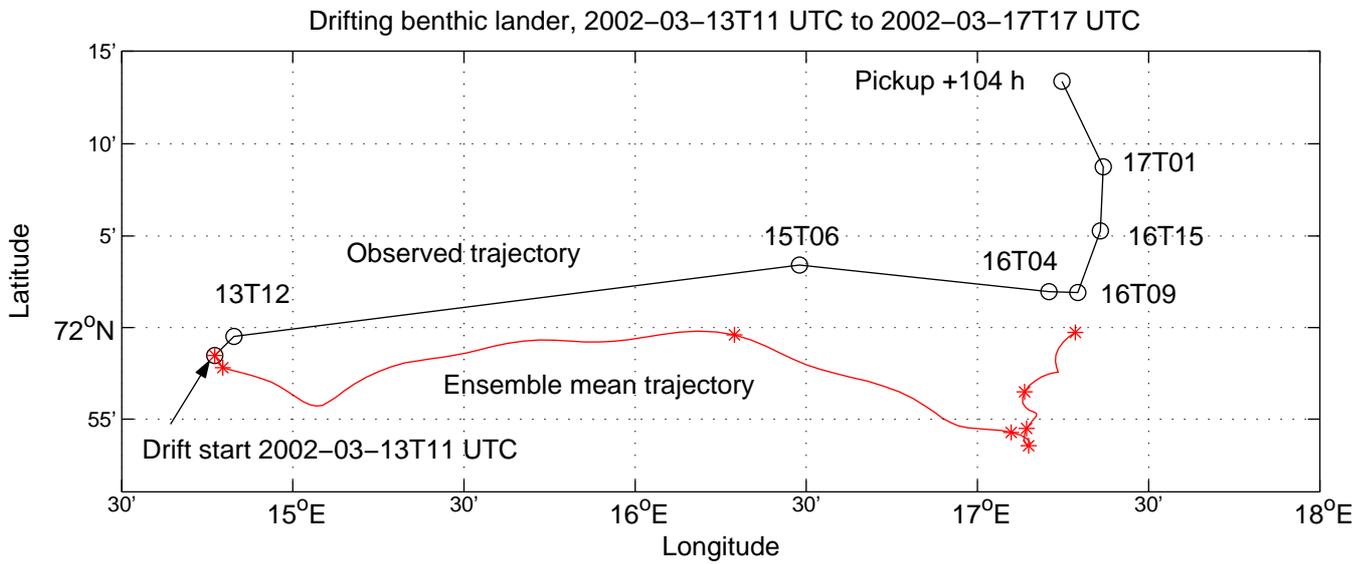}

\caption{\label{Fig:mod_v_obs}The trajectory of a benthic lander off the coast
of Northern Norway. ARGOS positions marked with (o) and date-time.
Lower curve represents the ensemble mean model trajectory. Intermediate
positions coincident with times of observations are marked with ({*}).
The lander drifted for more than four days over which the general
features of the trajectory were well captured by the model.}

\end{figure}

A benthic lander released itself from the sea bed on 2002-03-14T11
UTC after a malfunction. Its position was tracked irregularly (observations
are marked with date and time in Fig~\ref{Fig:mod_v_obs}) by the
ARGOS network over the following days until it was picked up on 2002-03-17T17
UTC. As Fig~\ref{Fig:mod_v_obs} shows, the average trajectory using
the class {}``person in water, mean values'' (PIW-1) corresponds
quite well with the observed intermediate positions (compare the asterisks
on the ensemble mean trajectory with the observed positions, marked
with date and time.

\subsubsection*{Faroe-Iceland exercise 2003}

\begin{figure}
\includegraphics{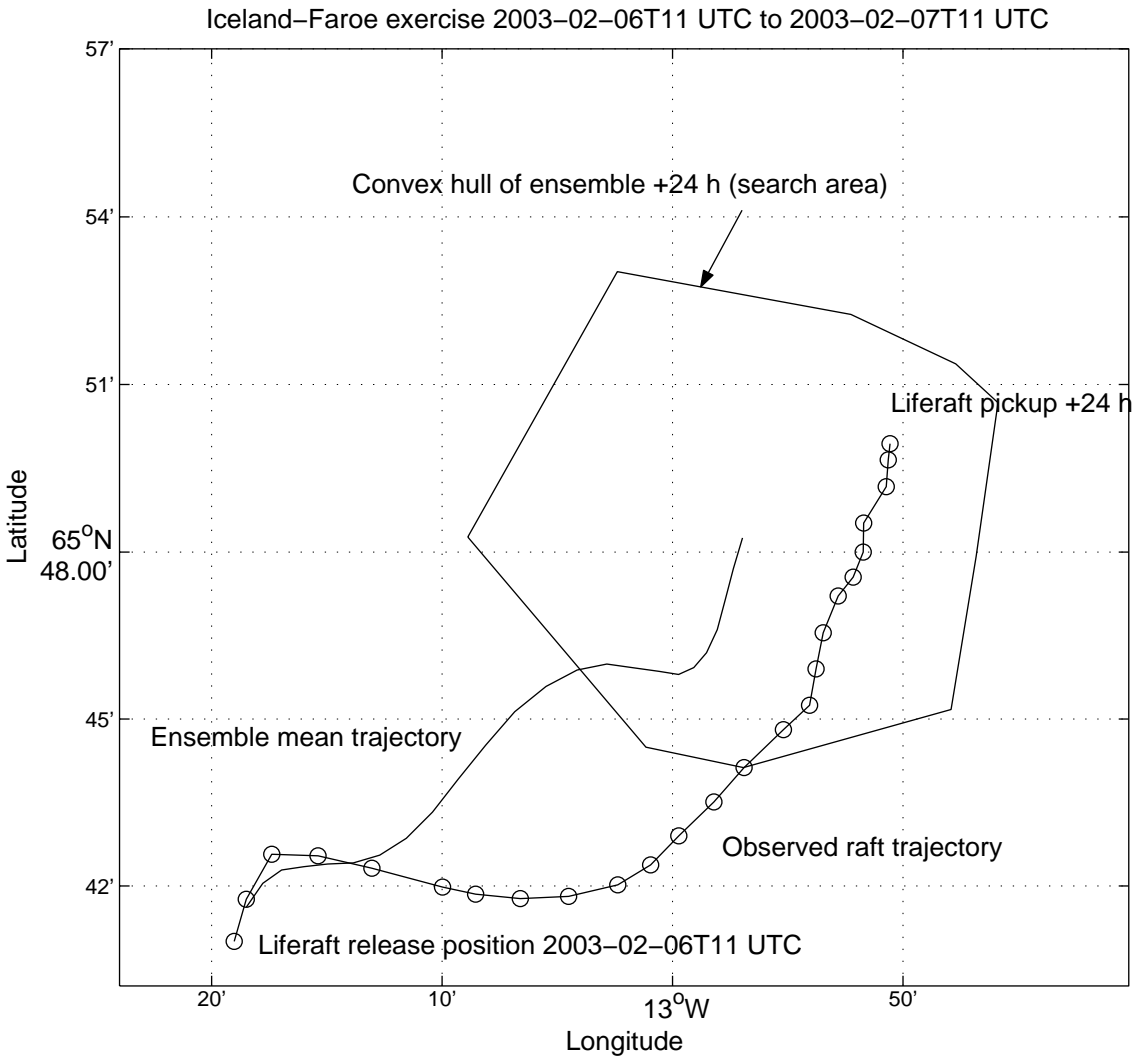}

\caption{\label{Fig:search_area_iceland}Average trajectory \emph{v} observations.
A liferaft was tracked for 24 hours, its GPS positions are marked
with (o). The ensemble mean model trajectory compared reasonably well
with the intermediate positions. The search ended after 24 hours.
By then the search area, illustrated by the convex hull of the ensemble,
had expanded to approximately 90 $\textrm{nm}^{2}$.}

\end{figure}

A liferaft was released on 2003-03-21T10 UTC as part of an exercise
conducted by the Faroese, Icelandic, and Norwegian rescue services.
Its position was tracked for 24 hours using GPS. As Fig~\ref{Fig:search_area_iceland}
shows, the liferaft was picked up near the center of the search area.
The ensemble mean trajectory agreed well with the intermediate positions.

\subsubsection*{Comparison of search methods: Faroe-Iceland exercise 2004}

\begin{figure}
\includegraphics{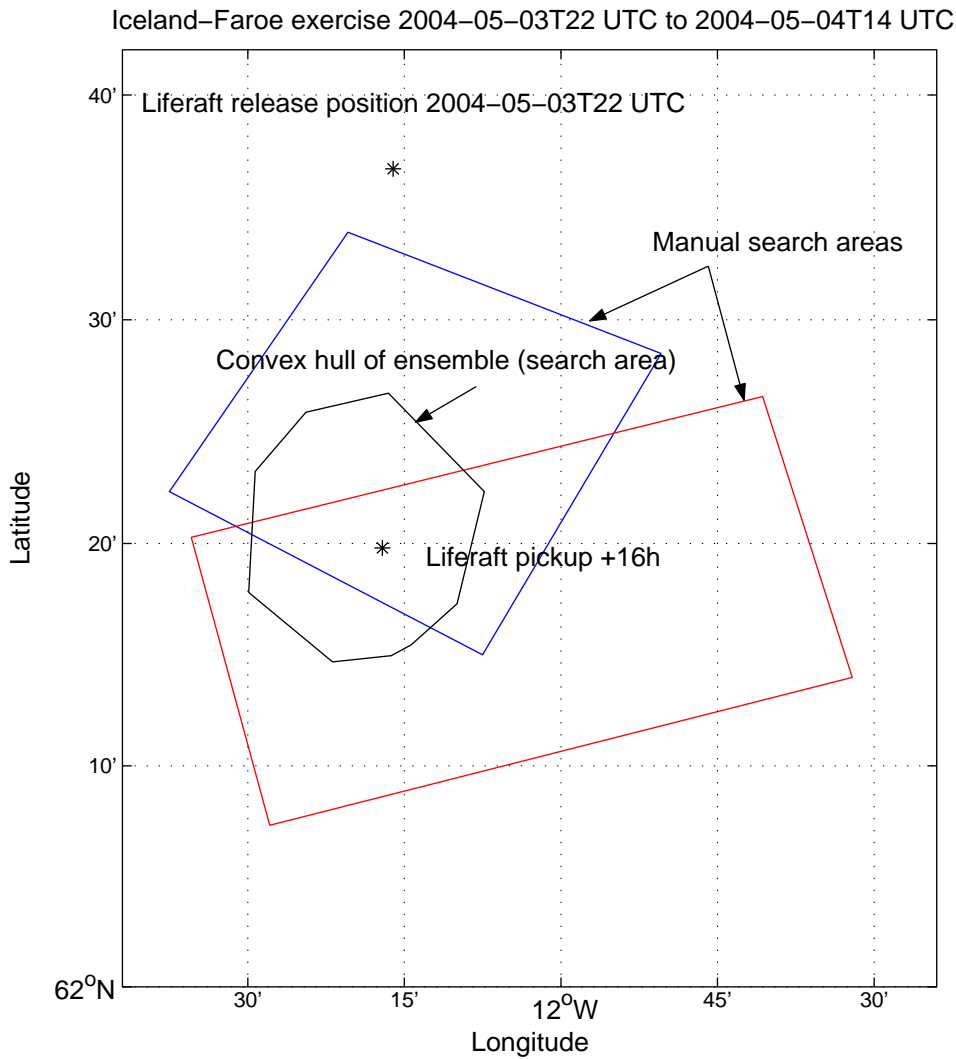}

\caption{\label{Fig:search_area_faroe}Comparison of search methods. Three
search methods were used to forecast the search area for a liferaft
16 hours after its release. The two search areas marked {}``manual
search areas'' were computed using the standard methods employed
by the Icelandic and Faroese SAR organizations. These methods rely
on climatological surface currents and low-resolution wind forecasts.
The search ended after 16 hours. By then the search area, illustrated
by the convex hull (the polygon) of the ensemble, had expanded to
100 $\textrm{nm}^{2}$, significantly smaller than the manual search
areas.}

\end{figure}

A liferaft was released on 2004-05-04T08 UTC through a joint exercise
between the Faroese, Icelandic, and Norwegian rescue services. The
pickup position and search areas computed using three different methods
are shown in Fig~\ref{Fig:search_area_faroe}. The search area based
on the \textsc{Leeway} model is marked with black. As can be seen
the raft is picked up near the centre of the search area. A comparison
with search areas computed by the Faroe and Icelandic rescue services
(red and blue) suggests that the rate of expansion of search areas
based on the \textsc{Leeway} model is 25-50\% of that using older
methods. No intermediate positions of the raft were available.

\subsection{Possible model extensions and improvements}

The \textsc{Leeway} model is built primarily for operational purposes.
As such it must be fast, easy to operate (limited degrees of freedom),
and it must not underestimate the rate of expansion of search areas.
Here we investigate possible extensions to the model which may have
an impact on the precision and the rate of expansion of the search
areas.

\subsubsection*{Improved object taxonomy and leeway coefficients}

The leeway categories available today are a compilation of all the
field experiments performed by the US Coast Guard, the Canadian Coast
Guard, and other search and rescue services around the world (see
Table~\ref{Tab:taxonomy}). The datasets are of various quality and
some object classes have a very high error variance (see the example
in Fig~\ref{Fig:raft_vs_sailboat}). Such classes have not been studied
with modern field equipment and should be redone. It is important
to note also that because the perturbations to the drift properties
are time-invariant, they contribute much more than the random walk
added to the wind field. Lowering the experimental variance of the
leeway coefficients for a search object translates directly into a
reduced the rate of inflation of the search area.

Several object classes are clearly missing as each country with a
coastline has its own particular SAR objects\emph{,} e.g., fishing
boats endemic to the region. There are also many world-wide SAR categories
that would benefit from further refinement, e.g., the various brands
of life rafts.

\subsubsection*{Jibing, swamping and capsizing}

The empirical trajectory model described in Sec~\ref{sec:theory}
computes the linear displacement of a SAR object by the wind and the
currents. In reality, a drifting object in rough weather will experience
both breaking waves and strong wind gusts. The object may jibe, swamp,
or capsize under such conditions. No attempt has been made at trying
to include these effects as experimental data are scarce. However,
with dedicated experiments it may be possible to quantify the probabilities
of jibing, swamping, and capsizing for different SAR objects.

Jibing has been observed to occur at very low wind speeds and at very
high wind speeds. It may be assumed that there is a small, but finite,
probability that an object will jibe under intermediate wind conditions
\cite{all05}. The effect of including jibes will be a more continuous
search area as some ensemble members will {}``change sides'' and
fill in the gap between the two distributions.

Swamping and capsizing affect the leeway of the object drastically.
Assigning probabilities of capsizing and swamping under different
weather conditions would contribute toward making the search area
more realistic. For now, lack of experimental data prevents a thorough
assessment of the frequency of swamping and capsizing for different
categories of SAR objects.

\subsubsection*{Increased ocean model resolution}

Although it is by no means certain that higher resolution leads to
higher precision in ocean modelling, increasing the horizontal resolution
is important for two reasons. 

First, even at 4 km resolution the model is only beginning to resolve
the eddy activity of the upper ocean. Moving towards 1-2 km resolution
will finally yield a realistic current spectrum. Resolving eddies
means that a cloud of particles (or ensemble) will disperse more realistically,
even if the eddies are not located correctly in time and space. 

\begin{figure}
\includegraphics[scale=0.4]{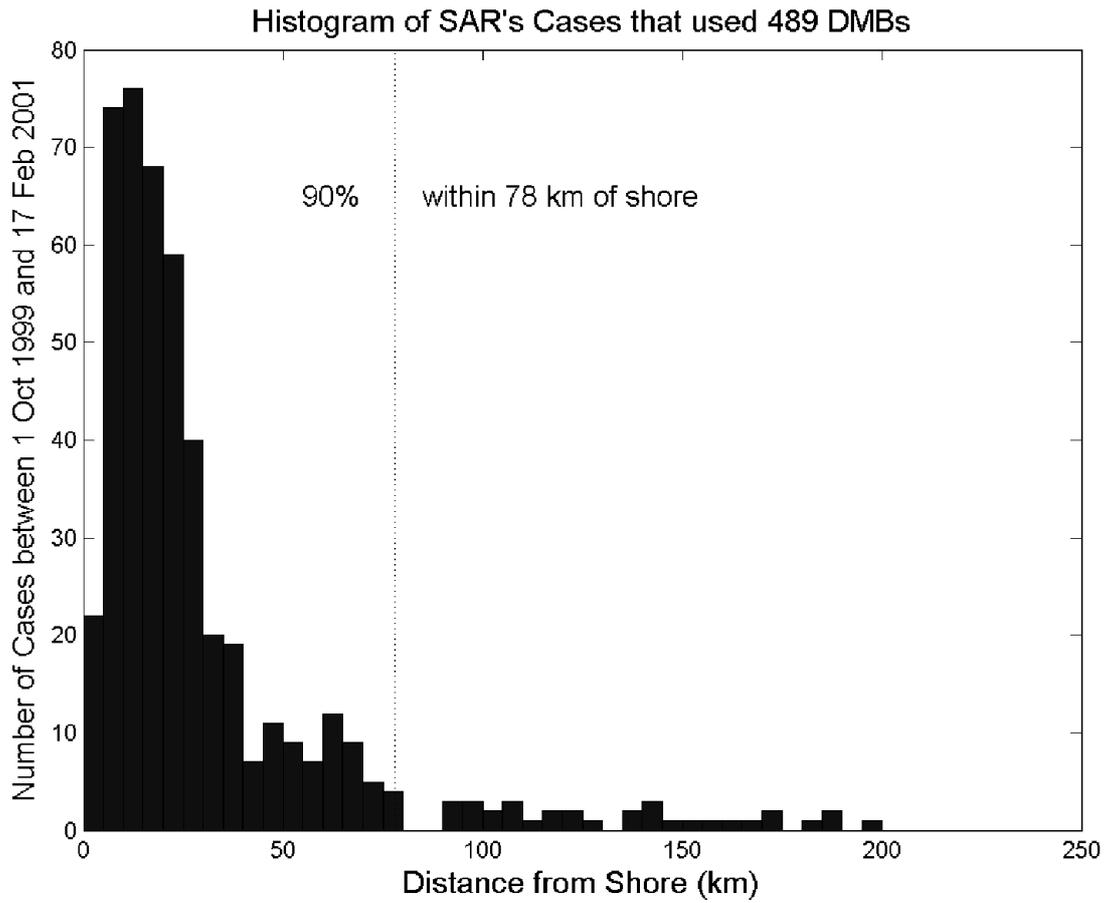}

\caption{\label{Fig:SARhist}Number of US Coast Guard SAR cases involving datum
marker buoys between 1999-10-01 and 2001-02-17 and their distance
to shore.}

\end{figure}

Second, the vast majority of rescue operations take place less than
40 km (25 nm) from shore. Shown in Fig~\ref{Fig:SARhist} is the
distribution of self-locating datum marker buoy (SLDMB) with distance
from shore in incidents in US waters. The buoys are deployed to assess
the speed and direction of the surface current near the last known
position of a search object and thus indicate the typical seaward
distance of incidents. Obviously, surface current fields that resolve
the near-shore currents well will greatly enhance the value of an
operational SAR model. For complex coastlines (the Norwegian is a
prime example) detailed current fields are required to realistically
model the flow between islands and in the mouths of fjords and estuaries.

\subsection{Conclusion}

We have presented a new operational model for the evolution of search
areas for drifting objects. A taxonomy of common SAR objects has been
set up based on the field experiments conducted to date. 

A new method for decomposing and perturbing the leeway of the object
in downwind and crosswind components has been employed, yielding more
robust computations at low winds and more realistic ensemble perturbations.
It is found that this new method makes search areas inflate at approximately
25-50\% of the rate found using older methods. The stochastic particle
trajectory approach employed for this model thus leads to a significantly
lower rate of expansion of search areas compared with other models.

The ensemble trajectory model is operational and can forecast search
areas up to 60 hours ahead in time. A seven day archive of wind and
current fields allows simulations to be started as early as one week
ago. This is important, as incidents are not always reported immediately.

It is found that particles are dispersed primarily through the perturbation
of leeway properties, as these remain constant for each individual
ensemble member. The added dispersion from perturbations (random walk)
of the wind field and the current field is negligible in comparison.
Going to a higher-order stochastic particle model is found to make
very little difference to the rate of expansion of search areas. Thus
a random walk model for the perturbations of wind and current fields
is sufficiently sophisticated to capture the evolution of search areas
given the high dispersion caused by the time-invariant leeway coefficients.

The vast majority of rescue operations at sea occur close to the coast.
We have argued that further increasing the horizontal resolution of
the ocean model will allow the model to assist searches even nearer
the shore and in major bays and fjords while at the same time making
the ensemble spread more realistically as eddy motion becomes more
adequately resolved.

More field work is required to adequately model the various SAR objects
frequently found in Norwegian waters and elsewhere. Furthermore, the
experimental error of some of the older leeway categories is so high
that it is desirable to revisit these SAR objects by conducting new
and improved field experiments. 

We have a limited, yet convincing set of field trials where the model
has been compared with trajectories and end positions of real SAR
objects (life rafts). The model succeeds in capturing the features
of the trajectories of several drifting objects. Further studies will
be required before definite conclusions can be drawn about its forecast
skill, but the results so far are promising.

\subsection*{Acknowledgments}
This work was supported by the Norwegian Ministry of Justice, The
Norwegian Joint Rescue Coordination Centres (JRCC), and The Royal
Norwegian Navy through the project ``Drift av gjenstander'' (Drifting
objects) and subsequent follow-up projects. The JRCC also contributed
directly to this work by making the data from their field exercises
available. The Research Council of Norway and the French-Norwegian
Foundation (Eureka grant E!3652) supported the development of the
operational service through the SAR-DRIFT project. The U.S. Coast
Guard has generously made all their compiled leeway field data available.
\bibliographystyle{breivik}
\bibliography{/home/oyvindb/Doc/TeX/Bibtex/BreivikAbb,/home/oyvindb/Doc/TeX/Bibtex/Breivik}

 \clearpage

\section*{Figures}



\clearpage

\section*{Tables}

\begin{table}
\caption{\label{Tab:taxonomy}Truncated SAR target taxonomy. The modern deep
ballast rafts are further broken down into subclasses for loading,
capacity, and the existence of a drogue (sea anchor). The leeway data
for these categories are compiled by Allen and Plourde (1999) from
field experiments performed by various search and rescue organizations
worldwide.}

\begin{tabular}{|l|l||l|l||l|}
\hline 
\multicolumn{5}{|l|}{Leeway target classes }\tabularnewline
\hline
Person  & \multicolumn{4}{l|}{Vertical }\tabularnewline
\cline{2-5} 
 in & \multicolumn{4}{l|}{Sitting }\tabularnewline
\cline{2-5} 
 Water & \multicolumn{2}{l|}{ } & \multicolumn{2}{l|}{Survival Suit }\tabularnewline
\cline{4-5} 
 (PIW) & \multicolumn{2}{l|}{Horizontal } & \multicolumn{2}{l|}{Scuba Suit }\tabularnewline
\cline{4-5} 
 & \multicolumn{2}{l|}{ } & \multicolumn{2}{l|}{Deceased }\tabularnewline
\hline
 & \multicolumn{2}{l|}{Maritime } & \multicolumn{2}{l|}{No Ballast System }\tabularnewline
\cline{4-5} 
 & \multicolumn{2}{l|}{Life } & \multicolumn{2}{l|}{Shallow Ballast System w/ Canopy }\tabularnewline
\cline{4-5} 
 Survival  & \multicolumn{2}{l|}{Rafts } & \multicolumn{2}{l|}{Deep Ballast System w/ Canopy }\tabularnewline
\cline{2-5} 
 Craft & \multicolumn{2}{l|}{Other } & \multicolumn{2}{l|}{Life Capsule }\tabularnewline
\cline{4-5} 
 & \multicolumn{2}{l|}{Maritime } & \multicolumn{2}{l|}{USCG Sea Rescue Kit }\tabularnewline
\cline{2-5} 
 & \multicolumn{2}{l|}{Aviation Life } & \multicolumn{2}{l|}{No Ballast w/ Canopy }\tabularnewline
\cline{4-5} 
 & \multicolumn{2}{l|}{ Rafts } & \multicolumn{2}{l|}{Evacuation Slide }\tabularnewline
\hline
Person & \multicolumn{4}{l|}{Sea Kayak }\tabularnewline
\cline{2-5} 
 Powered & \multicolumn{4}{l|}{Surf Board }\tabularnewline
\cline{2-5} 
 Craft & \multicolumn{4}{l|}{Windsurfer }\tabularnewline
\hline
Sailing  & \multicolumn{2}{l|}{Mono-hull } & \multicolumn{2}{l|}{Full Keel, Deep Draft }\tabularnewline
\cline{4-5} 
 Vessels & \multicolumn{2}{l|}{ } & \multicolumn{2}{l|}{Fin Keel, Shoal Draft }\tabularnewline
\hline
 & \multicolumn{2}{l|}{Skiffs } & \multicolumn{2}{l|}{Flat Bottom Boston Whaler }\tabularnewline
\cline{4-5} 
 & \multicolumn{2}{l|}{ } & \multicolumn{2}{l|}{V-hull }\tabularnewline
\cline{2-5} 
 & \multicolumn{2}{l|}{Sport Boat } & \multicolumn{2}{l|}{V-hull Cuddy Cabin }\tabularnewline
\cline{2-5} 
 Power  & \multicolumn{2}{l|}{Sport Fisher } & \multicolumn{2}{l|}{Center Console, Open Cockpit }\tabularnewline
\cline{2-5} 
 Vessels & \multicolumn{2}{l|}{Commercial } & \multicolumn{2}{l|}{Hawaiian Sampan }\tabularnewline
\cline{4-5} 
 & \multicolumn{2}{l|}{Fishing } & \multicolumn{2}{l|}{Japanese Stern-trawler }\tabularnewline
\cline{4-5} 
 & \multicolumn{2}{l|}{Vessels } & \multicolumn{2}{l|}{Japanese Longliners }\tabularnewline
\cline{4-5} 
 & \multicolumn{2}{l|}{(F/V) } & \multicolumn{2}{l|}{Korean F/V }\tabularnewline
\cline{4-5} 
 & \multicolumn{2}{l|}{ } & \multicolumn{2}{l|}{Gill-netter w/ Rear Reel }\tabularnewline
\hline
Boating & \multicolumn{4}{l|}{F/V Debris }\tabularnewline
\cline{2-5} 
 Debris & \multicolumn{4}{l|}{Bait / Wharf Box }\tabularnewline
\hline
Immigration  & \multicolumn{2}{l|}{Cuban Refugee} & \multicolumn{2}{l|}{With Sail }\tabularnewline
\cline{4-5} 
 Vessel & \multicolumn{2}{l|}{ Raft} & \multicolumn{2}{l|}{Without Sail }\tabularnewline
\hline
\end{tabular}
\end{table}

\end{document}